\documentclass[12pt,preprint]{aastex}
\def\fermi{{\it Fermi}}
\usepackage{amsmath}

\slugcomment{Not to appear in Nonlearned J., 45.}

\shorttitle{High-energy emissions from the Vela pulsar}
\shortauthors{Leung et al. }

\begin{document}


\title{Fermi-LAT Detection of Pulsed Gamma-rays Above 50 GeV \\from the Vela Pulsar}


\author{Leung, Gene C. K.\altaffilmark{1}, Takata, J.\altaffilmark{1}, Ng, C.W.\altaffilmark{1}}
\author{Kong, A.K.H\altaffilmark{2}., Tam, P.H.T.\altaffilmark{2}}
\author{Hui, C.Y.\altaffilmark{3}}
\and
\author{Cheng, K.S.\altaffilmark{1}}
\email{gene930@connect.hku.hk, takata@hku.hk}


\altaffiltext{1}{Department of Physics, The University of Hong Kong, Pokfulam Road, Hong Kong}
\altaffiltext{2}{Institute of Astronomy and Department of Physics, 
National Tsing Hua University, Hsinchu, Taiwan}
\altaffiltext{3}{Department of Astronomy and Space Science, 
Chungnam National University, Daejeon, Republic of Korea}


\begin{abstract}
The First \fermi-LAT Catalog of Sources Above 10 GeV reported evidence of pulsed
emission above 25 GeV from 12 pulsars, including the Vela pulsar, which showed
evidence of pulsation at  $>37$ GeV energy bands. 
Using 62 months of \fermi-LAT data, we analyzed the gamma-ray emission from the 
Vela pulsar and searched for pulsed emission above 50 GeV. 
Having confirmed the 
significance of the pulsation in 30-50 GeV with 
the H-test (p-value $\sim10^{-77}$),
we extracted its pulse profile using the Bayesian block algorithm 
and compared it with the distribution of the 5 observed photons above 
50~GeV using the likelihood ratio test. Pulsation was significantly detected 
for photons above 50~GeV with p-value $=3\times10^{-5}$ ($4.2\sigma$). 
The detection of pulsation is significant above $4\sigma$ at $>79$ GeV 
and above $3\sigma$ at $>90$ GeV energy bands, making this the highest
 energy pulsation significantly detected by the LAT.  We explore non-stationary 
 outer gap scenario of  the very high-energy emissions 
from the Vela pulsar.   
\end{abstract}





\section{Introduction}
The mechanism of the GeV gamma-ray emissions from pulsars 
 remains to be solved since the discovery of  
gamma-ray emitting pulsars (Crab and Vela) in the 1970s. 
In the past several years, the observations of 
pulsar emissions above 10 GeV by the \fermi~Large Area
 Telescope (\fermi-LAT) and ground based
 Cherenkov telescopes 
have created a breakthrough in the understanding of the nature of  
GeV gamma-ray emissions from  the Crab pulsar.  
In particular, the detection of the pulsed emission up to 400 GeV by 
the MAGIC collaboration  (Aliu et al. 2008; Aleksic et al. 2014) and 
the VERITAS collaboration (2011) were unexpected by 
the standard curvature radiation scenario of the outer magnetospheric gap.
The observed spectrum between 100 MeV and 400 GeV is 
better described by a broken power law than 
the power law with exponential cutoff in the standard curvature 
radiation scenario. This result suggests that the inverse-Compton 
scattering process produces the emissions above 10 GeV of the Crab pulsar 
(Aleksik et al. 2011; Aharonian et al. 2012). 

 In the First \fermi-LAT Catalog of Sources 
Above 10 GeV \citep[1FHL,][]{1fhl},
20 pulsars were found to show pulsed emissions (p-value$\le0.05$) 
in the energy range $>10$ GeV, including 12 pulsars show pulsed emissions
 at  $>25$ GeV \citep{sp2012}. 1FHL also reported  
that evidence of pulsation from the Vela pulsar at $>37$GeV and two other pulsars (PSRs J0614-3329 and J1954+2836) at $>60$ GeV.
 Recently, pulsed emission of the Vela pulsar above 30 GeV was also 
reported by H.E.S.S. collaboration (2014), with a mean photon energy of 40 GeV.

Given the high significance detection of the Vela pulsar in 1FHL 
(e.g. $TS = 72$ in 30-100 GeV), it is a good candidate for studying 
pulsed emission at high energy. The recent release of the reprocessed
 Pass 7 (P7REP) LAT data provides significant improvement over Pass 7 (P7) 
in the direction reconstruction of high energy photons $>3$ GeV as well as improvement in energy reconstruction \citep{bregeon13}. For instance, the $95\%$ ($68\%$) 
containment angle for 'Source' class events at $\sim50$ GeV is reduced by 
$\sim40\%$ from $0\fdg8$ ($0\fdg2$) in P7 to $0\fdg5$ ($0\fdg12$) in P7REP. 
The reduced PSF at high energy implies that the number of background photons, 
primarily from Vela-X in this case, inside 
the $95\%$ containment angle of the target source
is also reduced, leading to an enhanced signal-to-noise at high energy.

In this paper, we report on the analysis of 62 months of \fermi-LAT data and 
the high significance ($4.2\sigma$) detection of pulsed emissions at 
$>50$ GeV from the Vela pulsar. The detection of pulsation is
significant above $4\sigma$ at $>79$ GeV and above $3\sigma$ 
at $>90$ GeV energy bands.

\section{Observations and Data Analysis}

In this study, we analyzed \fermi-LAT data collected for 62 months, from 2008 August 4 
to 2013 October 18. The data was reduced and analyzed using the \fermi~Science Tools 
package (v9r32p5), available from the \fermi~Science Support Center \footnote{http://
fermi.gsfc.nasa.gov/ssc/data/analysis/software/}. We selected only events in the 
reprocessed Pass 7 'Source' class and used the P7REP\_SOURCE\_V15 IRFs. 
To reduce contamination from the Earth's albedo, we 
excluded time intervals when the zenith angle of the ROI was greater than 
$100\degr$ or the LAT's rocking angle was greater than 52\degr. We 
calculated the pulse phases using the \texttt{TEMPO2} plugin for \fermi, with the timing 
model of the Vela pulsar available from \fermi-LAT Multiwavelength Coordinating Group 
\footnote{https://confluence.slac.stanford.edu/display/GLAMCOG/LAT+Gamma-ray
+Pulsar+Timing+Models} \citep{ray11}, which covers the entire span of the observation.

\subsection{Spectral Analysis}
\label{sec:spec}

We performed binned likelihood analyses using the \texttt{gtlike} 
tool in the region surrounding the Vela pulsar. We selected photons between 0.1 and 300 
GeV within a $20\degr\times20\degr$ ROI centered at the position of the Vela pulsar. We 
modeled the diffuse backgrounds with gll\_iem\_v05\_rev1.fit and 
iso\_source\_v05.txt. We included 
2FGL sources \citep{nolan12} within $15\degr$ of the ROI center
in the model. Only the 
spectral parameters of sources within $8\degr$ of the ROI center were left free, 
while others were fixed to the catalog values. Since the publication of 2FGL, extended 
emission from two nearby SNRs, Vela-Jr \citep{tana11} and Puppis A \citep{hew12}, have 
been reported. The spatial and spectral models of the PWN Vela-X was also updated 
\citep{gron13}. To model these extended sources, we first performed an off-pulse 
analysis for the region. 
We selected photons in the phase interval 0.8-1.0 (cf. Figure \ref{edlc}) for the off-pulse 
analysis as in \citet{gron13}.

We replaced the 7 point sources in 2FGL that were possibly associated to Vela-Jr and 
Puppis A \citep{nolan12} with the best-fit uniform circular disk spatial templates and simple power laws reported 
in \citet{tana11} and \citet{hew12}. For Vela-X, we replaced the uniform circular disk 
model in 2FGL with the best-fit uniform elliptical disk template
and broken power law reported by \citet{gron13}. 
Only the flux normalization parameters were left free. The off-pulse analysis was 
done with the Vela pulsar removed. 

The off-pulse fit results, after correcting for exposure difference, were then passed to a 
phase-averaged fit. The 
fluxes of the 3 extended sources were fixed to the exposure-corrected off-pulse values. 
The Vela pulsar was modeled with a power law with exponential cutoff
\begin{equation}
\frac{dN}{dE} = N_0 \left(\frac{E}{E_0}\right)^{-\Gamma} \exp\left[-\left(\frac{E}{E_
\mathrm{cut}}\right)^b\right].
\label{fit}
\end{equation}
The best-fit values for the phase-averaged spectrum are $\Gamma=1.086\pm0.004$, $E_\mathrm{cut}=383\pm5$ MeV and $b=0.510\pm0.002$. The best fit function 
is over-plotted in Figure~\ref{spec}.

Spectral points were obtained by performing likelihood fits in individual energy bands, fitting only the normalization parameters of the Vela pulsar
and point sources within $4\degr$, modeled 
as power laws, and the diffuse backgrounds. All other sources 
were fixed to the best-fit full band values. For the energy bands 50-100 GeV and 100-300 
GeV, the normalization parameter of Vela-X was also left free. The results are 
shown in Figure \ref{spec}. The Vela pulsar is detected in 50-100 GeV with a TS value of 
9.5 ($3.1\sigma$), and a photon flux of $(3.17\pm1.78)\times10^{-11}$ 
ph cm$^{-2}$ s$^{-1}$. In 100-300 GeV, the TS value of the Vela pulsar drops to 2.4 
($1.5\sigma$), and the 95$\%$ c.l. upper limit was shown. 

\begin{figure}
\centering
\includegraphics[width=1.\textwidth]{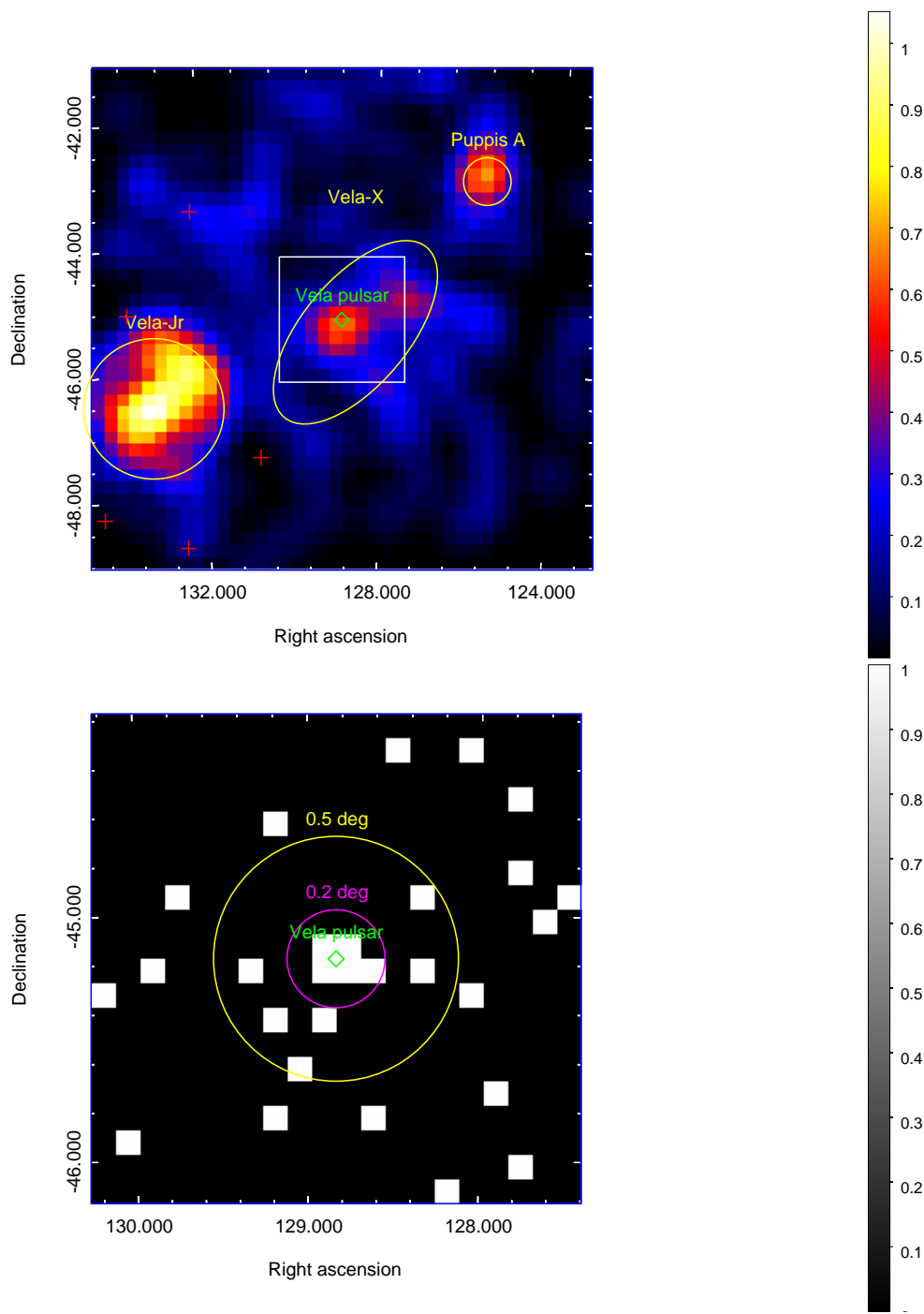}
\caption{Upper panel: Smoothed counts map of the $8\degr\times8\degr$ region with $0\fdg2\times0\fdg2$ pixel size. The Vela pulsar is labelled with a green diamond, 2FGL sources with red crosses and extended sources with yellow ellipses. The white square represents the region shown in the lower panel. Lower panel: Counts map of the $2\degr\times2\degr$ region with $0\fdg1\times0\fdg1$ pixel size. No smoothing is applied. The 5 photons with $P_{PSR} > \mathrm{max}(P_{PWN},P_{GAL})$ are all located within the $0\fdg2$ radius circle.}
\label{cmap}
\end{figure}

A spectral fit was performed in 50-300 GeV.
The source model was the same as above, except that the Vela pulsar and 
Vela-X were now modeled as power laws with
the normalization factors and spectral indices 
left free. The Vela pulsar was detected with a TS value of 11.7 ($3\sigma$),
with a photon index of $-2.53\pm0.98$ and a photon flux of 
$(4.27\pm2.13)\times10^{-11}$ ph cm$^{-2}$ s$^{-1}$.
Figure \ref{cmap} (upper panel) 
shows the smoothed counts map above 50 GeV in a $8\degr\times8\degr$ region. 
The emission around the position of the Vela pulsar is point-like and distinct from  
that of Vela-X. In contrast, the emission from Vela-Jr, with a spatial template of 
comparable size, is visibly extended.

\subsection{Pulsation Search}

\subsubsection{Weighted H-test}
\label{sec:wH}

Given the significant detection of the Vela pulsar in 50-300 GeV, we performed a 
pulsation search in this energy range. We applied a weighted 
H-test \citep{kerr11} to this energy range.
Each photon within $4\degr$ from the Vela pulsar was weighted by its probability
to have originated from the pulsar ($P_{PSR}$), calculated using 
the \texttt{gtsrcprob} tool.
We used the spectral model obtained from the spectral fit in 
50-300 GeV described in Section \ref{sec:spec} to calculate the probabilities.
The sum of probabilities $\sum{P_{PSR}}=6.3$, and the weighted
H-statistic is 15.4, corresponding to a p-value of $0.002$ or $3.1\sigma$.
Figure \ref{edlc}(a) shows the weighted light curve in 50-300 GeV.
The weighted H-test for higher energy ranges returned $H<15$, i.e. below $3\sigma$.

As noted in 1FHL, the H-test \citep{jag89,kerr11}
is not the most sensitive tool for the pulsation search on known pulsars at high energy, 
as it does not utilize any prior knowledge of the pulse shape. 
Therefore, we apply a likelihood ratio test in our search. 
The test compares the distribution in pulse phase of high energy photons with 
a known pulse profile in lower energies, and determines whether the high energy photons 
are better described by the known pulse profile or a uniform distribution. For this test to 
be most effective, it is necessary that the lower energy pulse profile should satisfactorily 
reflect the expected distribution of high energy photons.

\subsubsection{Selection of Low Energy Pulse Profile}

\begin{figure}
\centering
\includegraphics[width=1.\textwidth]{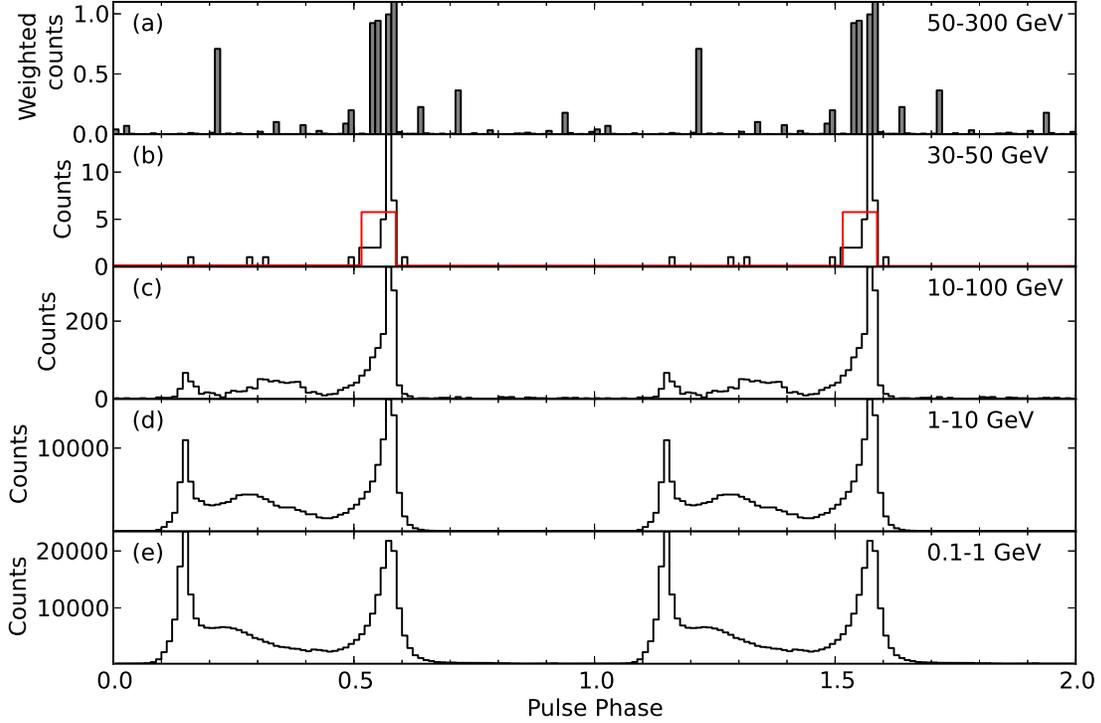}
\caption{ (a) Weighted light curve in 50-300 GeV. (b) Folded light curve in 30-50 GeV with a $0\fdg4$ radius aperture. The black histogram represents the observed counts, the red line represents the Bayesian block decomposition. Folded light curves in (c) 10-100 GeV, (d) 1-10 GeV and (e) 0.1-1 GeV with a $1\degr$ radius aperture. }
\label{edlc}
\end{figure}

As shown in previous studies \citep[e.g.][]{vela10, 2pc}, the pulse profile of the Vela 
pulsar is strongly dependent on energy. Figure \ref{edlc}(c) to (e) shows the folded light 
curves of the Vela pulsar in 3 energy bands: 0.1-1 GeV, 1-10 GeV and 10-100 GeV. With 
increasing energy, P1 weakens significantly, with the peak heights ratio P1:P2 
decreasing by a factor of $\sim5$ from 0.1-1 GeV to 10-100 GeV. The bridge structure 
also shifts towards P2 and reduces in height. Therefore, it is expected that the photons 
above 50 GeV should follow a distribution more similar to that in 10-100 GeV and 
concentrate around P2, and it is not appropriate to apply the full energy pulse profile in 
the likelihood ratio test.

In order to obtain a pulse profile that better reflects the high energy behavior, we need to 
use an energy range high enough to reflect the expected distribution of the $>50$ GeV 
photons, but low enough to demonstrate a statistically significant pulse profile.  

We extracted the folded light curve in 30-50 GeV and performed the unweighted 
H-test \citep{jag89}. We experimented with different aperture radii 
in steps of $0\fdg1$ and found that a $0\fdg4$ 
radius maximizes the H-statistic, giving $H=440$ (p-value $\sim10^{-77}$). 
This represents a statistically significant pulse 
profile to be used in the likelihood ratio test.

We then employed the Bayesian block algorithm \citep{sca13} to identify statistically 
significant variation features in the folded light curve in 30-50 GeV in a $0\fdg4$ 
radius aperture, using a false positive threshold of $0.5\%$. Since the Bayesian 
block algorithm assumes Poisson statistics,
it can only be applied to unweighted photons. 
The folded light curve in 30-50 GeV is represented by a block 
spanning the phase interval from 0.5159 to 0.5870. 
Figure \ref{edlc}(b) shows the 
30-50 GeV folded light curve overlaid with the Bayesian blocks. This block function was 
taken to be the low energy pulse profile in the likelihood ratio test.

\subsubsection{Photon Selection Above 50 GeV}

To select $>50$ GeV photons for the likelihood ratio test, we calculated also the 
probabilities of coming from Vela-X ($P_{PWN}$) and the Galactic diffuse 
emission ($P_{GAL}$) for photons within 
$4\degr$ from the Vela pulsar, following the procedures described in Section 
\ref{sec:wH}. Since the Vela pulsar is completely embedded in Vela-X 
as seen by the LAT, the major sources of background contamination are 
Vela-X and the Galactic diffuse emission. Therefore, we selected only photons 
with $P_{PSR} > \mathrm{max}(P_{PWN},P_{GAL})$.

\begin{figure}
\centering
\includegraphics[width=1.\textwidth]{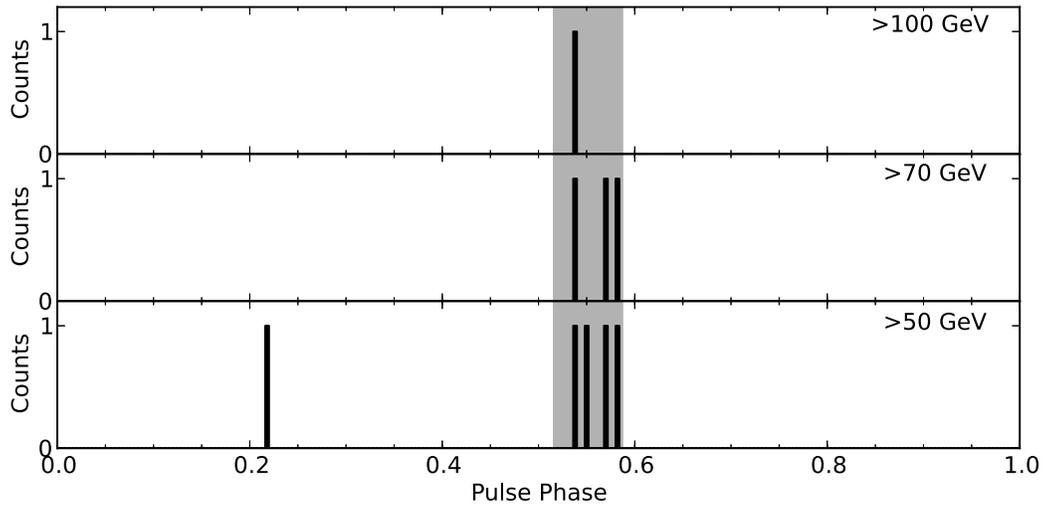}
\caption{Folded light curves for 50-300 GeV (bottom), 70-300 GeV (middle) and 100-300 GeV (top). The shaded area represents the on-peak interval decided by Bayesian blocks at 30-50 GeV.}
\label{lc50}
\end{figure}

A total of 5 photons have $P_{PSR} > \mathrm{max}(P_{PWN},P_{GAL})$. Figure 
\ref{cmap} (lower panel) displays the counts map above 50 GeV. All the 5 photons are 
located within the $0\fdg2$ circle centered at the Vela pulsar. We also note that all 5 
photons fall into the highest quality 'Ultraclean' class of events, meaning that they have 
the strictest rejection against cosmic ray events, which are more common 
at higher energies. 
Figure \ref{lc50} shows the folded light curves in 3 energy bands: 50-300 GeV, 70-300 
GeV and 100-300 GeV. The shaded area represents the phase interval spanned by the 
Bayesian block in 30-50 GeV. 4 of the 5 photons are located within the Bayesian block 
and are distributed across all 3 energy bands. The highest energy photon within the 
Bayesian block is at 209 GeV. Table \ref{tab1} lists the energy, arrival time, 
angular separation from 
the Vela pulsar, pulse phase and source probability of the 5 photons. We note 
that one glitch occurred at MJD 53959.9 \citep{yu13} before the start of the observation, 
and two other glitches were recorded during the span of the observation at MJD 55408.8 
and MJD 56555.8 in the timing model. None of the selected photons was detected 
near a glitch epoch.

 \subsubsection{Likelihood Ratio Test}
 
We performed a maximum likelihood fit to the $>50$ GeV folded light curve using a 
probability distribution of a block function
\begin{equation}
PDF(\phi) = 
\left\{
	\begin{array}{ll}
	1-s+\dfrac{s}{\phi_1-\phi_0} & \mathrm{if}~\phi_0\le\phi\le\phi_1\\
	1-s & \mathrm{otherwise,}
	\end{array}
\right.
\end{equation}
where $\phi$ is the pulse phase, $\phi_0$ and $\phi_1$ are fixed to the edges of 
the Bayesian block in 30-50 GeV, and $s$ is a free parameter in [0,1].
We also performed a fit with the null hypothesis of a uniform 
distribution, i.e. $s=0$. Using the likelihood ratio test, we obtained the test statistic, 
$D = -2\Delta\log{(\mathrm{Likelihood})} = 16.3$ for $E>50$ GeV.
We scanned up in energy by removing the lowest-energy photon each time 
and calculating the $D$ value. The $D$ values are listed in Table
\ref{tab1}.

Since the Wilk's theorem, which translates $D$ to p-values,
assumes infinite statistics,
we calculated the p-values with Monte Carlo simulations. In each Monte 
Carlo realization, $N$ phases, where $N$ is drawn from a Poisson 
distribution with the mean equal to the observed number of photons, 
were randomly generated in $[0,1]$. Then the same analysis was performed 
to the simulated data to calculate the $D$ value. 
The p-values from Monte Carlo simulation agree with
Wilk's theorem, and are listed in Table \ref{tab1}.
The significance of pulsation is above $4\sigma$ for $>51$ and $>79$ GeV,
and above $3\sigma$ at  $>90$ GeV energy bands.
 We note that the pulsation significance
for $>55$ GeV is below $4\sigma$, due to the contribution of an off-pulse photon
at 55.9 GeV. We also note that, however, this particular photon has a smaller 
probability ($\sim0.7$) to have originated from the Vela pulsar than the other four photons ($>0.9$).

\section{Discussion}
\begin{figure}
\centering
\includegraphics[width=1.\textwidth]{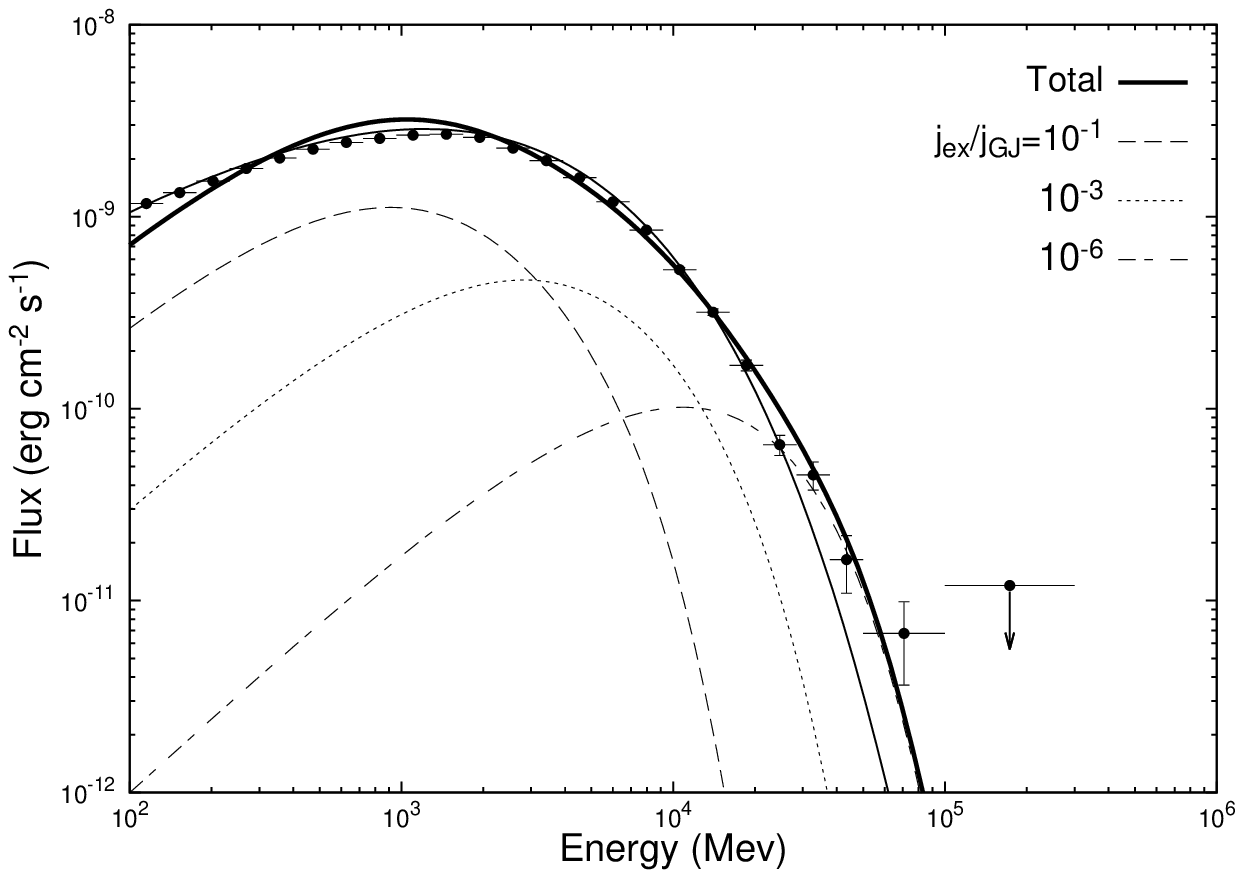}
\caption{Phase-averaged gamma-ray spectrum of the Vela pulsar. The thick solid 
line shows the spectrum superposed the emissions from the 
 outer gap structures with the different injection currents. 
The result is for power index  $p=0.6$ 
as the distribution of the external current (equation~\ref{dist}). 
The dashed, dotted and dashed-dotted lines show  individual 
contribution of the outer gap with $j_{ex}/j_{GJ}=10^{-1}$, 
$10^{-3}$ and $10^{-6}$, respectively. We assume the inclination angle
 $\alpha=65$degree. The thin solid line shows the observational fit 
described by equation of (\ref{fit}).}
\label{spec}
\end{figure}

As shown in Figure~\ref{spec}, the observed flux above 10GeV of the 
Vela pulsar  decreases  slower than a simple exponential function. 
Previous emission models that invoke the curvature radiation process 
for the GeV emissions have  in general predicted a flux at the 50-100GeV 
smaller than  the observed flux level of $10^{-12}{\rm erg cm^{-2}s^{-1}}$. 
 Abdo et al. (2010) proposed that a sub-exponential cut-off 
in the observed spectrum can be understood as 
the superposition of several power low plus exponential cut-off functions with 
varying the photon index and the cut-off energy. 

In the section, we argue the model that the outer gap accelerator 
 is switching between a number of states, and that superposition of 
the emissions from the different states of the outer gap make 
the observed spectrum of $Fermi$.  There is a wide range of variability 
times scale in the radio emissions 
from the pulsar \citep[e.g.][]{kra02, ly10, ke13}. 
The micro-second variations seen in single pulse  
could be produced by spatial fluctuation in the emission region.
 The pulse-to-pulse variations on the timescale 
of millisecond to second likely represent timescale of 
 the temporal variation of the 
structure of the emission region. The longer timescale (second to year)
 variations associated with the  model switching and nulling, 
which sometimes accompany the variations of 
the spin down rate,  could be related with the changes  
of entire magnetosphere. The higher energy observations also found 
the mode switches in the X-ray emission properties 
of PSR B0943+10 \citep{he13} 
and in the GeV gamma-ray emission properties of PSR J2021+4026 \citep{al13}. 
These multi-wavelength observations suggest that the switching 
between a number of state of magnetosphere is probably 
a generic feature of the pulsars.

The active  outer gap should require  
the external currents ($j_{ex}$) that are injected into the outer
 gap at the boundaries, because they initiate the gamma-ray emissions 
and subsequent pair-creation cascade process. In our model, 
the outer gap structure
 and the properties of the  curvature radiation depend on how large 
currents are injected into the gap. For example, the 
 outer gap size can develop until the pair-creation process creates 
the gap current  of order of the Goldreich-Julian
 value $j_{GJ}\sim \Omega B/2\pi$ \citep{takata04, hirotani06}.  
The outer gap with  a smaller external  current has 
 a larger gap size and a larger electric field, and  produces  
 harder  gamma-rays. If the external current, which 
may be originated from the polar cap region, is temporally  variable, 
the observed gamma-ray spectrum will be  the superposition 
of the emissions from various gap states.  

We consider that the outer gap 
 \citep{cheng86, takata06} produces the GeV emissions; 
in the model the charge depletion from the Goldreich-Julian 
charge density causes the electric field along 
the magnetic field line, which accelerates the electrons and positrons
 to Lorentz factor of $\sim 10^7$.  We assume the dipole magnetic field in the 
magnetosphere and solve three-dimensional structure of 
the gap (Takata et al. in prepare); (1) the particle acceleration 
process by the electric field, (2) the curvature radiation process 
and (3) the pair-creation process 
between the gamma-rays and thermal X-rays from the neutron star surface. 
Our 3-D model is the expansion of 2-D study developed by 
Takata et al. (2006; 2008; 2009), which discussed the phase-averaged spectra 
of the pulsars measured by the EGRET.

In our model, we assume that the observed gamma-ray spectrum are superposition 
of the emissions  from various {\it stationary} gap structures with various 
external currents. The model assumes that the timescale of the variability 
of the external current is of order of or longer 
than crossing time scale of the light 
cylinder radius ($r_{lc}$), $\tau\sim r_{lc}/c= p/2\pi$, 
with which the stationary outer gap structure 
for an external current is archived.  Figure~\ref{spec} compares 
the calculated spectra with the observations. In the calcualtion, 
we assumed the power law 
distribution of the external current  as 
\begin{equation}
I=K j_{ex}^{p},~~j_{min}<j_{ex}<j_{max}
\label{dist}
\end{equation}
where $K$ is the normalization factor and is determined from 
$\int (dI/dj_{ex})dj_{ex}=1$, and  we used 
$j_{min}=10^{-6}j_{GJ}$ and  $j_{max}=0.2j_{GJ}$. 
In addition, we chose the power index $p\sim0.6$, 
which reasonably reproduces the observed spectral cut-off behavior. 
In Figure~\ref{spec}, the calculated 
spectrum for   {\it an external current} cannot explain the 
observed spectral shape in 100MeV-50GeV, 
while the gamma-ray spectrum superposed by the different 
gap states explains  better the observed  cut-off behavior. 
We also find that the emissions above 20GeV
 can be explained by the outer gap with a very small injection current.
 In the subsequent paper, we will study dependency of the viewing 
geometry (inclination angle and viewing angle) 
on the predicted spectrum and will discuss how goodness 
of the model fitting for high-energy pulsars.

In summary, we detected pulsed gamma-ray emissions from the 
Vela pulsar at above $4\sigma$ at  $>79$ GeV and above $3\sigma$ at  
$>90$ GeV energy bands using \fermi-LAT. This is difficult to explain using
the previous stationary model in the pulsar magnetosphere. We proposed the 
model that the outer gap structure is switching between a number of state.
Future searches for pulsed emission above 
50 GeV in other pulsars will help understand the nature of the high-energy 
emission in the pulsar magnetosphere.

\acknowledgments

We thank anonymous referee, Saz Parkinson, P.M. and McEnery, J.E. 
for useful comments and suggestions, and Wu, E.M.H. and Wu, J.H.K. 
for providing the weighted
H-test computing code. This work is partially supported by a 2014 GRF grant 
under HKU 17300814P and Seed Funds under HKU 20127159004 and 201310159026.
 AKHK and PHT are supported by the National Science Council of the
Republic of China (Taiwan) through grants 100-2628-M-007-002-MY3 
and 100-2923-M-007-001-MY3, and 101-2112-M-007-022-MY3 respectively. 
CYH is supported by a research fund from Chungnam National University in 2014.
 JT thanks TIARA operated under the ASIAA of Taiwan to use their PC cluster.

\clearpage





\begin{table}
\begin{center}
\begin{tabular}{ r | r r r r | r r r}
\hline \hline
$E_\gamma$ (GeV) & Time (MJD) & $\Delta\theta$ ($\degr$) & Pulse Phase & $P_{PSR}$ & $D_{E\ge E_\gamma}$ & $p_{E\ge E_\gamma}$ & $\sigma_{E\ge E_\gamma}$\\
\hline
51.3 & 55050.8 & 0.074 & 0.548 & 0.940 & 16.3 & $3.3\times10^{-5}$ & $4.2\sigma$\\
55.9 & 56149.3 & 0.136 & 0.219 & 0.707 & 11.5 & $4.2\times10^{-4}$ & $3.5\sigma$\\
79.5 & 56317.2 & 0.034 & 0.581 & 0.986 & 15.9 & $6.0\times10^{-5}$ & $4.0\sigma$\\
91.0 & 56437.5 & 0.011 & 0.569 & 0.994 & 10.6 & $1.4\times10^{-3}$ & $3.3\sigma$\\
208.5 & 55154.1 & 0.092 & 0.537 & 0.922 & 5.3 & $2.8\times10^{-2}$ & $2.2\sigma$\\
\hline \hline
\end{tabular} \\
\caption{Energy, arrival time, angular separation from the Vela pulsar, pulse phase and 
source probability of $>50$ GeV photons with $P_{PSR} > \mathrm{max}
(P_{PWN},P_{GAL})$ in ascending order of energy. Three glitches occurred at MJD 
53959.9, MJD 55408.8 and MJD 56555.8. The $D$ value, p-value and significance 
of pulsation at the energy range bounded below by $E_\gamma$ of each photon are 
also shown.}
\label{tab1}
\end{center}
\end{table}


\end{document}